\documentclass[12pt]{article}

\usepackage{latexsym}
\usepackage{epsfig}
\usepackage{psfrag}

\def\void{}
\def\labelmark{}

\newenvironment{formula}[1]{\def\labelname{#1}
\ifx\void\labelname\def\junk{\begin{displaymath}}
\else\def\junk{\begin{equation}\label{\labelname}}\fi\junk}%
{\ifx\void\labelname\def\junk{\end{displaymath}}
\else\def\junk{\end{equation}}\fi\junk\labelmark\def\labelname{}}

{\ifx\void\labelname\def\junk{\end{array}\end{displaymath}}
\else\def\junk{\end{array}\right.\end{equation}}
\fi\junk\labelmark\def\labelname{}\def\junk{}
}

\newcommand{\beq}{\begin{formula}}
\newcommand{\eeq}{\end{formula}}
\newcommand{\beqv}{\begin{formula}{}}

\newcommand{\rf}[1]{(\ref{#1})}
\newcommand{\oh}{\frac{1}{2}}

\newcommand{\bea}{\begin{eqnarray}}
\newcommand{\eea}{\end{eqnarray}}
\newcommand{\beas}{\begin{eqnarray*}}
\newcommand{\eeas}{\end{eqnarray*}}
\newcommand{\beqs}{\begin{displaymath}}
\newcommand{\eeqs}{\end{displaymath}}

\newcommand{\br}{\langle}
\newcommand{\kt}{\rangle}

\newcommand{\ep}{\varepsilon}

\newcommand{\vp}{\varphi}

\newcommand{\ben}{\begin{equation}}
\newcommand{\een}{\end{equation}}

\newcommand{\bdm}{\begin{displaymath}}
\newcommand{\edm}{\end{displaymath}}

\newcommand{\pa}{\partial}

\newcommand{\bR}{{\bf R}}

\newcommand{\bon}{{\bf n}}

\begin{document}
 \topmargin 0pt
 \oddsidemargin 5mm
 \headheight 0pt
 \topskip 0mm

 \addtolength{\baselineskip}{0.4\baselineskip}

 \pagestyle{empty}

 \vspace{0.5cm}

\hfill NORDITA 2000/101 HE 

\hfill NBI-HE-00-44

\vspace{2cm}

\begin{center}

{\Large \bf Noncommutative Scalar Solitons:}

\medskip

{\Large\bf Existence and Nonexistence}



\vspace{1.2 truecm}



 \vspace{0.7 truecm}
Bergfinnur Durhuus$^a$\footnote{email: durhuus@math.ku.dk}, Thordur 
Jonsson$^b$\footnote{e-mail: thjons@raunvis.hi.is}\footnote{
Permanent address: Science Institute, University of Iceland, 
Dunhaga 3,  107 Reykjavik, 
Iceland} and Ryszard Nest$^a$\footnote{email: rnest@math.ku.dk}

\vspace{1 truecm}

$^a$Matematisk Institut, Universitetsparken 5

2100 Copenhagen \O, Denmark

 \vspace{.8 truecm}

$^b$The Niels Bohr Institute and NORDITA, Blegdamsvej 17

2100 Copenhagen \O, Denmark

 \vspace{1.5 truecm}

 \vspace{.8 truecm}

 \end{center}

 \noindent
 {\bf Abstract.} We study the variational equations for solitons in
noncommutative scalar field theories in an even number of spatial dimensions.
We prove the existence of spherically symmetric solutions for a
sufficiently large noncommutativity parameter $\theta$ 
and we prove the absence of spherically symmetric 
solutions for small $\theta$.

 \vfill

 \newpage
 \pagestyle{plain}

 \section{Introduction}  
Recently there has been considerable interest in
solitons in noncommutative field theories, the main
motivation coming from string theory \cite{connes,witten}.   
Several authors have found explicit solitons in gauge theories
with and without matter fields \cite{poly,aganagic,bak,gross1,harvey}.
   
In \cite{gsm1} solitons in scalar field theories were studied and 
it was shown  that there is a host of solitons 
in the case of an infinite
noncommutativity parameter $\theta$ 
where the kinetic term in the action can be neglected.
This is in a stark contrast to the commutative case where no
solitons exist \cite{derrick}.
The authors of \cite{gsm1} 
conjectured the existence of solitonic solutions at large $\theta$ and
indicated how to calculate them perturbatively in $\theta^{-1}$.  
In \cite{zhou} the existence problem for solitons at finte $\theta$ was 
discussed and it was argued that solitons would not exist for small
$\theta$.   Various aspects of solitons in noncommutative scalar
field theories are discussed in 
\cite{gorsky,lindstrom,solovyov,bak2,matsuo}.

In this paper we establish the existence of spherically symmetric solitons
in even dimensional scalar field theories under fairly general conditions
on the potential, provided $\theta$ is sufficiently large.  
We show that
no spherically symmetric solutions can exist for small $\theta$.  
We briefly comment on the problem of generalizing these results to the
nonrotationally invariant case.  In the bulk of the paper we deal, for 
simplicity, 
with the two-dimensional case and describe in the
last section the changes needed to treat the general even dimensional case.
				       
 \section{The problem} 
Solitons in a noncommutative two-dimensional scalar field theory with 
potential $V$  are defined as finite energy solutions to the variational 
equations of the action functional
\beq{1}
S(\vp )=\int \left(\oh (\nabla \vp )^2+V_\theta (\vp )\right)\, d^2x.
\eeq
Here the potential $V$ applied to the function $\varphi$ should be
calculated using the Moyal product $\star$ 
on the space of functions
on $\bR ^2$.  
If $V$ is a polynomial we can express $V_\theta (\vp )$ as
\beq{2}
V_\theta (\vp )=\sum_{j=1}^nc_j \vp ^{\star j}
\eeq
where the star powers are given by the star product
\beq{3}
(\vp \star\psi )(x)=\left.\exp\left(i{\theta\over 2} 
\ep_{jk}{\pa \over \pa y_j} 
{\pa \over \pa z_k}\right)\vp (y)\psi (z)\right|_{z=y=x}
\eeq
and we sum over repeated indices.
Here $\ep_{jk}$ is the antisymmetric symbol,
$\ep_{12}=-\ep_{21}=1$,  and $\theta >0$ is the 
noncommutativity parameter.  
We assume that $V(0)=0$ and $V(x) >0$ if $x\neq 0$.

The algebra of smooth rapidly decaying 
functions on $\bR ^2$ with the Moyal product 
defined above is well known to be isomorphic to an
algebra of operators on a one particle Hilbert space.  We denote this 
isomorphism by $\vp\mapsto \hat\vp$. Under the isomorphism
the $\star$-product becomes the usual operator product and
\beq{4}
{1\over 2\pi}
\int \vp (x)\,d^2x= \theta {\rm Tr}\hat\vp.
\eeq
For a discussion of this correspodence, see, e.g., \cite{gsm1,taylor}.  In the 
operator formalism the action functional becomes
\beq{5}
S (\hat \vp )={\rm Tr}\left( [a,\hat \vp][\hat\vp ,a^*]+\theta V(\hat \vp 
)\right),
\eeq
where $a^*$ and $a$ are the usual raising and lowering operators of the
simple 
harmonic oscillator and the operator $\hat\vp$ has been 
assumed to be self-adjoint which corresponds to a real valued function 
$\varphi$.
The variational equation of \rf{5} is 
\beq{6}
2[a,[a^*,\hat\vp ]]+\theta V'(\hat\vp )=0.
\eeq
If we choose the harmonic oscillator eigenstates as the basis for our 
Hilbert space then it is easily checked that radially symmetric functions
$\vp$ correspond to diagonal operators $\hat\vp$.  
If we consider a diagonal operator with eigenvalues $\lambda_n$, $n=0,1,2,
\ldots $, Eq.\ \rf{6}
reduces to \cite{gsm1,zhou}
\bea
(n+1)\lambda_{n+1}-(2n+1)\lambda_n +n\lambda_{n-1} & = & 
{\theta\over 2} V'(\lambda _n), ~~n \geq 1 \label{em}\\
\lambda_1-\lambda_0 & = & {\theta\over 2} V'(\lambda_0 ).\label{em0}
\eea
  We can sum the second order finite
difference equation for $\lambda_n$ from $n=0$ to $n=N$ and obtain
the first order equation
\beq{7}
\lambda_{N+1}-\lambda_N={\theta\over 2(N+1)}\sum_{n=0}^NV'(\lambda _n),
~~N\geq 0.
\eeq
A necessary condition for the energy to be finite is clearly that
\beq{bc}
\lambda_N\to 0 ~~\mbox{\rm  as}~~ N\to\infty.
\eeq
The first problem we wish to address 
is when there do exist solutions to Eq.\ \rf{7}
satisfying the boundary condition \rf{bc}.

\section{Properties of solutions}
In this section we derive some properties that any solution to
Eqs.\ \rf{7} and \rf{bc} must satsfy under some simple conditions on the 
potential $V$.  We assume that $V$ is twice 
continuously differentiable and has only one local minimum in addition to $0$.
Let the other local minum be at $s>0$.  
Let $r\in (0,s)$ be a point where $V$ has a local maximum and for technical
convenience assume that $V'$ does not vanish except at $0,r$ and $s$.
Then $V'(x)<0$ for $x<0$ or $x\in (r,s)$ and $V'(x)>0$ for $x>s$ or
$x\in (0,r)$.

Let us now assume that $\lambda_n$ is a sequence of real numbers which 
solves Eq.\ \rf{7} with the boundary condition \rf{bc}.  We assume that 
the $\lambda_n$'s are
not all zero.  In this case we
have the following:
\begin{itemize}
\item[(a)] $0< \lambda_n < s$, for all $n$.
\item[(b)] $\lambda_n$ tends monotonically to $0$ for $n$ large enough
\item[(c)] $\sum_nV'(\lambda _n)=0$ and  $\sum_n\lambda_n<\infty$.
\end{itemize}
We now establish these three properties in turn.  

Suppose $\lambda_0\geq s$.  Then $V'(\lambda_0)\geq 0$ so by Eq.\ \rf{7}
$\lambda_1\geq\lambda_0$ and $\lambda_n$ is a nondecreasing 
sequence by induction. 
Similarly,
if $\lambda_0<0$ then $\lambda_n$ is a decreasing sequence.  Hence, $0\leq 
\lambda_0< s$.  If $\lambda_n\notin (0,s)$ for some values of $n$, let
$p>0$ be the smallest such $n$.  
If $\lambda_p\geq s$ then $\lambda_p-\lambda_{p-1}>0$
and by Eq.\ \rf{7} it follows that 
$V'(\lambda_p)>0$.  By the previous argument we conclude that the sequence
$\lambda_n$ is increasing for $n>p$.  Similarly, if $\lambda_p\leq 0$ then 
the sequnce is decreasing for $n>p$.   In both cases the boundary condition
\rf{bc} is violated and this proves (a).  

The above argument shows that if $\lambda_{p+1}-\lambda_p>0$ and $0<
\lambda_p<r$ then $\lambda_n$ will be increasing for $n\geq p$, at least until
$\lambda_n>r$.  Hence, since $\lambda_n\to 0$, it follows that
$\lambda_{n+1}-\lambda_n\leq 0$ for $n$ large enough which establishes (b).
This was also noted in \cite{zhou}.

From (b) and \rf{7} we see that 
\beq{8}
S_N\equiv \sum_{n=1}^NV'(\lambda_n)\leq 0
\eeq
for $N$ large enough.  Since $\lambda_n\in (0,r]$ for $n$ large we 
conclude that the sum $S_N$ is increasing with $N$ for $N$ sufficiently
large.  Hence, 
\beq{9}
S_N\to S_\infty\equiv\sum_{n=0}^\infty V'(\lambda_n)\leq 0
\eeq 
as $N\to \infty$.  If $S_\infty=u<0$ for some $u$ then it follows from
eq.\ \rf{7} that
\beq{10}
\lambda_{n+1}-\lambda_n\sim {u\over 2\theta }{1\over n}
\eeq
for $n$ large.  It follows that
\beq{11}
\lambda_{p}-\lambda_{q}=\sum_{n=q}^{p-1}(\lambda_{n+1}-\lambda_n)\sim 
{u\over 2\theta}\sum_{n=q}^{p-1}{1\over n}\to -\infty
\eeq
for $p\to\infty$ contradicting $\lambda_p\to 0$ as $p\to \infty$.
We conclude that $S_\infty=0$ which also follows (formally) from the 
equation of motion by taking trace.

Since $V'$ is approximately linear and increasing on a neighbourhood of 
$0$ we conclude that the sums $S_\infty$ and
$
\sum_{n}\lambda_n
$
are absolutely convergent and the operator $\hat\vp$ is trace class.

We remark that if $\hat\vp$ is any solution to Eq.\ \rf{5} then one can
show by methods different from those of this paper 
that the spectrum of $\hat\vp$ is
contained in the interval $[0,s]$.  This property will presumably be 
helpful for the study of non-spherically-symmetric solutions to
Eq.\ \rf{5}.

\section{Existence}

In this section we use the results of the previous section and some 
elementary mathematical techniques to prove the existence of solutions to
Eqs.\ \rf{7} and \rf{bc}.  Let $t$ be the location of the maximum of 
$V'$ in the interval $[0,s]$ and let $w$ be the location of the 
minimum of $V'$ in the same interval.

\begin{figure}[thb]
  \begin{center}
    \psfrag{V'(x)}{$V'(x)$}
   \psfrag{x}{$x$}
  \psfrag{r}{$r$}
 \psfrag{s}{$s$}
    \includegraphics[width=10cm]{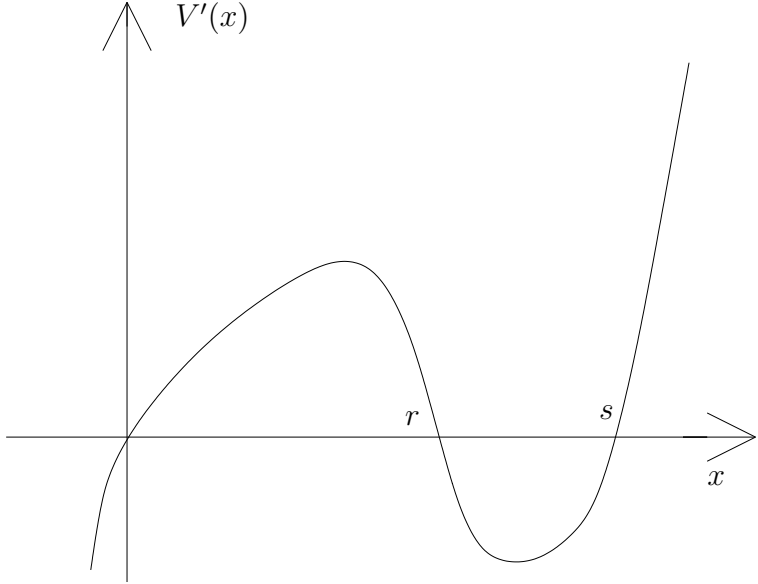}
    \caption{An graph of the derivative of a generic potential $V$
which satisfies our assumptions.}
    \label{fig1}
  \end{center}
\end{figure}

Let us begin by assuming that 
\beq{13}
\theta \geq -{2w\over V'(w)}.
\eeq
In this case there is a unique largest 
$\underline{\lambda}\in [w,s)$ such that if 
we set $\lambda_0=\underline{\lambda}$ in the recursion \rf{7} then
$\lambda_1=0$ and consequently 
$\lambda_n\to -\infty $ as $n\to\infty$.
Evidently $\underline{\lambda}$ increases monotonically to $s$ as 
$\theta\to\infty$.  Now take $s>\lambda_0> \underline{\lambda}$.  Then
$\lambda_1>0$.  Assuming that $\theta$ is large enough we have
\beq{14}
|V'(\underline{\lambda})|\leq V'(t)
\eeq
and it follows that there is a unique $\bar\lambda\in 
(\underline{\lambda},s)$ such that
\beq{15}
V'(\lambda_1)=-V'(\lambda_0)
\eeq
if $\lambda_0=\bar\lambda$ and
\beq{16}
V'(\lambda_1)<-V'(\bar\lambda)
\eeq
for $\lambda_0\in [\underline{\lambda}, \bar\lambda)$.
This means that for the sequence $\lambda_n$ defined by $\lambda_0=
\bar\lambda$ and the recursion \rf{7} we have 
$
0<\lambda_1=\lambda_2<\lambda_3$.
  If a sequence $\lambda_n$
obeys the recursion \rf{7} and has the property
$\lambda_0>\lambda_1>\ldots > \lambda_N$ but $\lambda_{N+1}\geq \lambda_N$
we say that the sequence {\em turns at} $N$.  
Note that in this case $\lambda_N>0$ by the proof of 
property (a) in Sec.\ 3.  
Furthermore, if $\lambda_{N+1}=\lambda_N$ then $\lambda_{N+2}>\lambda_{N+1}$.

Let us define the set
\beq{18}
A=\{\lambda_0\in [\underline{\lambda},\bar\lambda ]:
\lambda_n ~~\mbox{\rm turns at some}~~ N\}.
\eeq
By construction $\underline{\lambda}\notin A$ and $\bar\lambda\in A$.
Put $\Lambda_0=\inf A$.  Since each $\lambda_n$ depends continuously
on the initial value $\lambda_0$ it follows that $\Lambda_0<\bar\lambda$
and $\Lambda_0\notin A$.

Now consider the sequence defined by $\lambda_0=\Lambda_0$ and
Eq. \rf{7}.  Since this
sequence does not turn it is monotonically decreasing.  In order to
show that this sequence provides 
a solution to our problem it therefore suffices to 
show that the sequence converges to $0$.  
Suppose the sequence becomes negative for some $n$ and let $n=N$ be
the smallest value of $n$ such that $\lambda_N<0$.  Then by the
proof of property (a) in Sec.\ 3 we find that $\lambda_n\to -\infty$.
By the continuity of $\lambda_n$ as a function of $\lambda_0$ it 
follows that for $\lambda_0$ sufficiently close to $\Lambda_0$
the sequence $\lambda_n$ 
converges monotonically to $-\infty$  but this contradicts the definition 
of $\Lambda_0$ as the infimum of those $\lambda_0$-values for which 
the sequence turns.  
It thus follows that the limit $\lim_{n\to\infty}\lambda_n=a\geq$ exists
and by Eq.\ \rf{7} we have
\beq{19}
V'(a)={2\over \theta}\lim_{n\to\infty}(\lambda_{n+1}-\lambda_n)=0.
\eeq
Hence, $a=0$ since $\lambda_n\leq\lambda_1<r$ by construction.
This completes the proof of the existence of solutions for large $\theta$.

The solutions whose existence was established above converge in the limit
$\theta\to\infty$ to the projector on the ground state of the
harmonic oscillator, corresponding to $\lambda_0=s$ and $\lambda_n=0$
for $n\geq 1$.  We can easily generalize the above construction to obtain
different spherically symmetric solutions.  For 
example, we can tune the initial value $\lambda_0$ so that 
$\lambda_0,\lambda_1,\ldots ,\lambda_n\in (r,s)$ but $\lambda_{n+1}\in
(0,r)$ and the sequence is decreasing as before.  In the infinite $\theta$
limit this solution converges to the projector on the subspace 
spanned by the first $n$ harmonic oscillator states.  Working
slightly harder one can construct sequences with $\lambda_0$ close to
$0$ which jump to the interval $[r,s]$ at some $N$ and thereafter converge
monotonically to zero.  These solutions correspond to a projection on the 
$N$th eigenstate of the harmonic oscillator.

\section{Nonexistence}
In this section we will show that for sufficiently small $\theta$ 
the kinetic terms in the equation of motion, 
i.e., the left hand side of Eq.\ 
\rf{em} cannot match the potential terms on the right hand side.
Assume that the sequence $\lambda_n$ is a solution to Eqs.\ \rf{7} and \rf{bc}.
If we multiply Eq.\ \rf{em} by $\lambda_{n+1}-\lambda_{n-1}$ it can be 
written as
\bea
&&
(n+1)(\lambda_{n+1}-\lambda_n)^2-n(\lambda_n-\lambda_{n-1})^2
+(\lambda_{n+1}-\lambda_n)(\lambda_n-\lambda_{n-1}) \nonumber\\
&&={\theta\over 2}V'(\lambda_n)(\lambda_{n+1}-\lambda_n)+
{\theta\over 2}V'(\lambda_n)(\lambda_{n}-\lambda_{n-1})\label{30}.
\eea
Let us first assume that $\lambda_n$ is a decreasing 
sequence (as is the case for the solutions constructed in the previous 
section).  Then we sum Eq.\ \rf{30} over $n$ from $1$ to $\infty$.  
Defining $\Delta\lambda_n=(\lambda_{n+1}-\lambda_n)$ 
the result can be written
\beq{31}
-\Delta\lambda_0^2+\sum_{n=1}^\infty\Delta\lambda_n\Delta\lambda_{n-1}=
{\theta\over 2}\sum_{n=1}^\infty V'(\lambda_n)(\Delta\lambda_n+
\Delta\lambda_{n-1}).
\eeq
We note by Eq.\ \rf{7} and property (a) in Sec.\ 3 that
\beq{x1}
|\Delta\lambda_n|\leq c\theta
\eeq
where the constant $c$ depends only on the potential $V$ but 
is independent of $\theta$.  It follows that
the right hand side of Eq.\ \rf{31} can be written as
\beq{31x}
\theta\int_{\lambda_0}^0 V'(\lambda )\,d\lambda +O(\theta^2)=
-\theta V(\lambda_0)+O(\theta^2).
\eeq
We have 
\beq{32}
\sum_{n=1}^\infty\Delta\lambda_n\Delta\lambda_{n-1}\geq 0
\eeq
so Eq.\ \rf{31} implies
\beq{33}
0\leq -\theta V(\lambda_0)+O(\theta^2).
\eeq
By property (c) of solutions to Eq.\ \rf{7} we know that $\lambda_0>r$. 
It follows
that the inequality \rf{33} cannot be valid for small $\theta$ and we have a
contradiction.

If the eigenvalues $\lambda_n$ do not form a monotonic sequence we use 
property (b) of solutions and let $N$ be such that $\lambda_n \leq
\lambda_{n-1}$ for $n\geq N$ but $\lambda_{N-1}>\lambda_{N-2}$.  
Then $V'(\lambda_{N-1})\leq 0$ so $\lambda_{N-1}\geq r$.
Then we
can repeat the argument above by summing Eq.\ \rf{30} over $n$ from $N$
to $\infty$ and noting that
\beq{34}
|\lambda_N-\lambda_{N-1}|\leq{\theta\over 2 N}|V'(\lambda_{N-1})|.
\eeq
This completes the proof of the fact that for sufficiently 
small $\theta$ (depending on the potential $V$) there do
not exist any solutions to  Eqs.\ \rf{7} and \rf{bc}.

\section{Higher dimensions}

If space has dimension $2d$, $d>1$, we can choose coordinates such that 
the antisymmetric commutativity matrix is
of a $2$ by $2$ block diagonal form 
where each block is identical to the noncommutativity 
matrix $\theta\ep_{ij}$ in two dimensions.
The operators corresponding to the functions on $\bR ^{2d}$ now act on 
a Hilbert space which can be regarded as the Hilbert space of $d$ independent 
simple harmonic oscillators with raising and lowering operators
$a_i,a_i^*$, $i=1,\ldots d$.  As a basis in this space we choose the vectors
\beq{35x}
|\bon\kt =|n_1\ldots n_d\kt
\eeq
which are joint eigenvectors of the number operators 
$a_i^*a_i$ with nonnegative integer eigenvalues $n_i$, $i=1
\ldots d$.  It is readily checked that rotationally invariant functions
on $\bR ^{2d}$ correspond to operators of the form
\beq{36x}
\hat\vp =\sum_{\bon}\lambda_n|\bon\kt\br\bon |
\eeq
where the eigenvalues $\lambda_n$ only depend on
\beq{37x}
n\equiv \sum_{i=1}^dn_i.
\eeq
The equation of motion is now 
\beq{38x}
2\sum_{i=1}^d[a_i,[a^*_i,\hat\vp ]]+\theta V'(\hat\vp )=0.
\eeq
For operators of the form \rf{36x} we obtain from Eq.\ \rf{38x} 
the following genralization
of Eq.\ \rf{em}
\beq{39x}
(n+d)\lambda_{n+1}-(2n+d)\lambda_n+n\lambda_{n-1}={\theta\over 2}V'(\lambda_n).
\eeq
In order to derive a first order finite difference equation for the 
eigenvalues it is convenient to multiply Eq.\ \rf{39x} by a positive
coefficient $\alpha_n$ chosen such that when we sum the equation 
over $n$ from $0$ to
$N$ all the $\lambda_n$'s cancel except for $\lambda_N$ and $\lambda_{N-1}$.  
In order for this to happen we must have
\beq{40x}
\alpha_{n+1}={n+d\over n+1}\alpha_n
\eeq
so if we choose $\alpha_0=1$ then
\beq{41x}
\alpha_n={(n+d)\ldots (n+2)\over (d-1)!}.
\eeq
The generalization of Eq.\ \rf{7} is therefore
\beq{42x}
\lambda_{N+1}-\lambda_N={\theta\over 2(N+1)\alpha_N}\sum_{n=1}^N\alpha_n
V'(\lambda_n).
\eeq
For this recursion formula all the arguments of Secs.\ 3 and 4 go through
with minor modifications.  The same applies to the nonexistence proof for
small values of $\theta$ which is now obtained by summing Eq.\ \rf{39x},
multiplied on both sides by $\lambda_{n+1}-\lambda_{n-1}$, 
over $n$ from
the last turning point $N$ to $\infty$ and observing that inequalities
$|\Delta\lambda_n|\leq c\theta$ and $|\Delta\lambda_{N-1}|\leq c\theta 
|V'(\lambda_{N-1})|/N$ are still valid.

\section{Discussion}
In this paper we have established the existence of spherically symmetric
solitons in all sufficiently strongly 
noncommutative even dimensional scalar field theories with
potentials which increase at infinity and have two unequal minima.
Clearly these solutions can be translated to generate other solutions.
The existence and/or nonexistence (depending on $\theta$) 
of solitons which are not rotationally invariant about
any point is still an open problem.

The noncommutativity is responsible for the existence of 
these solitons which do not exist in the
corresponding commutative or weakly noncommutative theories.  The
assumption of spherical symmetry simplifies the analysis but is presumably 
not necessary to prove nonexistence at small $\theta$.  
At the present time we can generalize the nonexistence theorem of 
Derrick \cite{derrick} to the
weakly noncommutative setting under some regularity assumptions.  
This as well as the stability of solutions 
will be treated in a forthcoming paper.

\bigskip

\noindent
{\bf Acknowledgements} 
The work of B.D. is supported in
part by MatPhySto funded by the Danish National Research Foundation.  
This research was partly supported by TMR grant no. HPRN-CT-1999-00161.

\end{document}